**Nanowire design by magnetic collection of Fe, Ni and/or FeNi-alloy nanoparticles**

S. Ekeroth[1], R. D. Boyd[1], N. Brenning[1,2], U. Helmersson[1]

[1] Department of Physics, Chemistry and Biology, Linköping University, SE-581 83 Linköping, Sweden.

[2] KTH Royal Institute of Technology, School of Electrical Engineering, Division of Space and Plasma Physics, SE-100 44 Stockholm, Sweden.

*Abstract*

A method for growing nanoparticles with different elemental compositions simply by changing pulsing parameters in a power supply is demonstrated. The technique is based on high power pulsed hollow cathode sputtering, and the difference in particle composition ranges from pure Fe and Ni, to fully alloyed FeNi particles. Several pulses, or bursts, from a hollow cathode are needed in order to grow a nanoparticle using high power pulses. With this in mind, we devised a setup with two hollow cathodes of different material, Fe and Ni, situated in the vicinity of each other. By using two power supplies and a synchronization unit, the number and size of pulses of each element into the growth regime of the nanoparticles is controlled, and a tunable mixture of ions of the two elements is ensured. We here show that the two-target high power hollow cathode sputtering unit can be a powerful tool for rapid prototyping of more advanced, customized, nanoparticles, which can be of great importance for numerous applications.

*Introduction*

The ability to design nanoparticles and nanostructures to successfully fulfill the need in different applications are of great interest in many different research fields. We have previously demonstrated that magnetic nanoparticles can be generated from a magnetic material by the use of hollow-cathode sputtering where after the generated nanoparticles are self-assembled in a magnetic field to form nanowires and nanotrusses that are of great interest as electrodes, e.g. in electrocatalytic applications. [AAA] However, the intelligent design of suitable structures for optimized performance requires a larger degree of freedom in the design of the individual nanoparticles and how these nanoparticles are assembled. In view of this need, we here demonstrate the possibility to tune the elemental ratios within a nanoparticle by using two hollow cathodes, one Fe and one Ni. In this way, it is possible to design a range of different nanoparticles, going from pure Fe or Ni nanoparticles, Fe/Ni-alloyed particles, to core-shell particles. By alternating the conditions, by will, it is possible to generate a large amounts of different complex nanostructures, where the nanoparticle building block is altered in different ways. Thanks to the flexibility of the process, rapid prototyping opens up for screening of nanostructures for optimal performance in applications such as catalytically active electrodes.

*Experimental*

The experiments are performed using a pulsed hollow cathode method suitable for nanoparticle synthesis from a generated metal ion flux. Here it has been modified by placing two hollow cathodes in close vicinity to each other as schematically demonstrated in Fig. 1. The plasma in the two cathodes can be individual controlled to generate metal ion plumes, in the synthesis region just outside of the cathodes, of a predetermined composition. This composition is easily modulated in time by altering pulsing sequences and power levels supplied to the cathodes. Argon gas (purity : 99.9997%) is used as a process gas and is flown (120 sccm) through the cathodes out in to the synthesis region. The pumping speed is throttled to create a pressure of 0.8 torr in the vacuum chamber. To ensure efficient nucleation of nanoparticles, a low background level of oxygen is maintained in the system. This is obtained by constantly flowing 0.5 sccm of an $O_2$ gas diluted to 95% with Ar, giving an effective $O_2$ flow of 0.025 sccm. The hollow cathode plasmas are ignited using two separate HiPSTER 1 units from Ionautics AB, fed by dc power supplies MDX-1K from Advanced Energy. To set the pulse sequences to the two systems, a HiPSTER Sync unit from Ionautics AB is used. The nanoparticles are collected onto 10x10 mm Si wafers, coated with 200 nm of Ti, using a magnetic field. On the substrate the nanoparticles form interlinked nanowires, or nanotrusses, as has been demonstrated earlier. [AAA] SEM images, including EDX intensity measurements, were taken using a LEO 1550 Gemini. The TEM images were taken using a FEI Tecnat G2 operated at 200 kV. High-angle angular dark field (HAADF) and energy dispersive X-ray spectroscopy (EDX) are combined with scanning-TEM (STEM) analyses using an annular detector spanning an angular range from 80 to 260 mrad. The samples were here ultrasonicated in isopropanol and TEM grids with an amorphous carbon support were dipped in the solution and left to dry. Grazing incidence X-ray diffraction (GIXRD) measurements are performed using an Empyrean diffractometer in a parallel beam configuration with a line focused copper anode source (Cu Kα, l = 0.154 nm), operating at 45 kV and 40 mA. The primary beam is conditioned using a parallel beam mirror and a 1/4° divergence slit and in the secondary beam path a 0.27° parallel plate collimator is used. A PIXcel-3D detector is used as an open detector for the data acquisition. The grazing incidence X-ray diffraction scans are performed at an incidence angle of 1° in a 2θ range of 30–80° using a step size of 0.015° and a data collection time of 0.88 s/step.

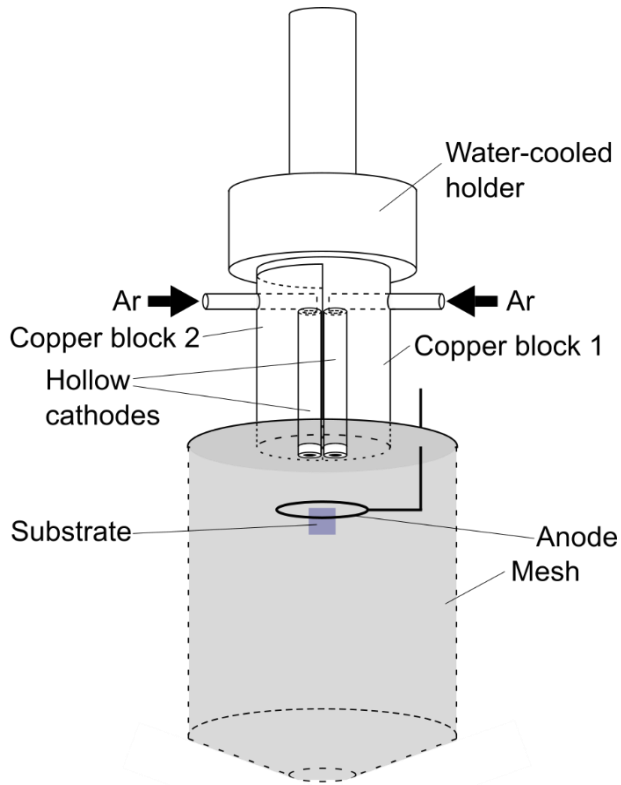
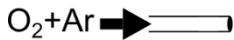

Fig. 1 | Schematic illustration of the hollow cathode setup.

*Results and Discussion*

Two groups of experiments are demonstrated. In the first group, the two cathodes are pulsed in a sequence with one pulse in one of the cathodes immediately followed with a pulse in the second cathode. This is repeated in a way that the pulsing frequency in each cathode is maintained at 600 Hz. This means that the pulsing frequency at the growth zone is kept at 1200 Hz (material independent). By adjusting the amount of power fed to the two different cathodes (sum of power always kept at 60 W) the metal ion composition in the synthesis zone can be altered, generating alloyed nanoparticles of a predetermined composition. Table 1 displays the set of pulse conditions used throughout the experiments within this paper. Each condition is given an ID number for easier reference throughout the paper, where i-v represents the group of experiments described above. A schematic illustration of the pulse delivery can be found in the supplementary information (SI), Fig. S1.

Table 1 | Pulse conditions with given ID number.

| Pulsing condition "ID" | Fe cathode power [W] | Ni cathode power [W] | Number of pulses in each Fe pulse train | Number of pulses in each Ni pulse train |
|---|---|---|---|---|
| i | 0 | 60 | 1 | 1 |
| ii | 15 | 45 | 1 | 1 |
| iii | 30 | 30 | 1 | 1 |
| iv | 45 | 15 | 1 | 1 |
| v | 60 | 0 | 1 | 1 |
| vi | 0 | 60 | 0 | 18000 |
| vii | 60 | 60 | 12000 | 24000 |
| viii | 60 | 60 | 18000 | 18000 |
| ix | 60 | 60 | 24000 | 12000 |
| x | 60 | 0 | 18000 | 0 |

To analyse the overall elemental composition of the nanoparticles within the samples in condition i-v, EDX analysis (in the SEM) is performed and the relative EDX intensities, $I_{Fe}/(I_{Fe}+I_{Ni})$, are plotted in Fig. 2 against the applied cathode power ratios, $P_{Fe}/(P_{Fe}+P_{Ni})$. The results illustrates a good correlation between applied power ratio and composition. This also implies that the desired Fe/Ni-ratio is achievable with relative ease simply by regulating the power. Of course, the power ratios needed to achieve different elemental ratios heavily depends on the parameters used; both for the source (e.g. sputter gas, power and pulse conditions) and for the sputter materials (e.g. sputter yield and secondary emission yield). However, these parameters are easy to investigate and customize for each material composition. For the parameters and materials used in this paper, the correlation between power-ratio and composition-ratio seems to be close to one-to-one.

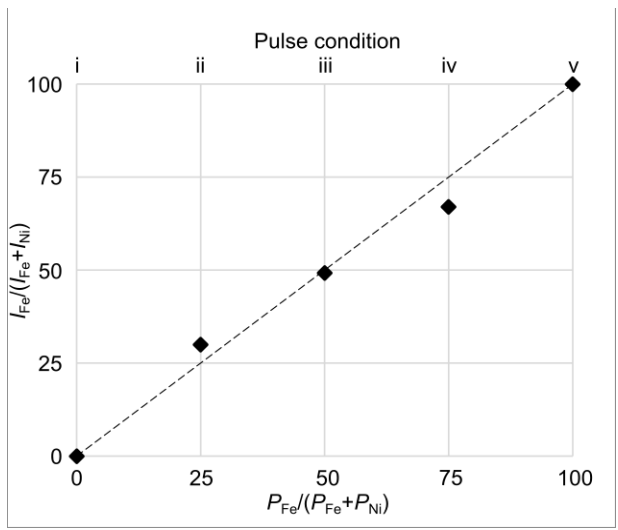

Fig. 2 | Correlation of the nanoparticles grown in condition i-v, in terms of EDX-peak intensity, to the fraction of power to the Fe cathode. The dashed trendline indicates a one-to-one correlation between power-ratio and composition-ratio. The samples corresponds to pulse condition i-v in Table 1.

HAADF of nanoparticles synthesized using the condition iii, along with compositional EDX-mapping of some of the particles, are presented in Fig. 3. The grown nanoparticles display the nanotruss formation typical for magnetic collection.[AAA] The figure show a homogeneous distribution of Fe and Ni in the particles, within the lateral resolution of the EDX. No apparent variation in-between nanoparticles are observed for a specific sample. The fact that these samples contain nanoparticles with a mixed Fe-Ni composition, instead of a mixture of single-species Fe and Ni nanoparticles, demonstrates that the formation of a nanoparticle takes longer time than the time of a single pulse. This has also been discussed in previous works, where the overlap of growth material from subsequent pulses was suggested to occur at higher pulse repetition frequencies.[BBB,CCC] This even distribution of Fe and Ni within the nanoparticles displayed in Fig. 3 is representative also for the conditions ii-iv, even though the elemental ratios are of course different for conditions ii and iv.

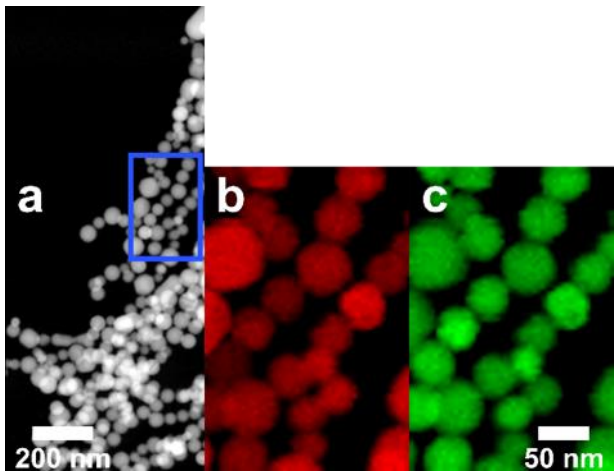

Fig. 3 | a) Dark field TEM image of a nanotruss grown using condition iii, collected into nanotrusses. b) and c) are EDX mapping of the area marked in a), of Fe and Ni, respectively.

In the second group of experiments, many pulses are fed to one cathode and after a short interruption, pulsing is initiated in the other cathode. This is repeated. In this way, the ion composition in the synthesis zone is stable over a time long enough to form nanoparticles of the same composition as of the particular cathode. This will generate nanoparticles of two different compositions, in the present experiment of pure Fe and pure Ni, that are used to synthesis nanowires/trusses of alternating nanoparticles. This group of experiments is marked vi-x in Table 1. Note that condition i and vi both yields only Ni nanoparticles as that is the only cathode ignited within these conditions. The only difference is that in condition vi the deposition is interrupted in order to mimic the rest of the pulse train conditions. For the same reason, condition v and x only yields Fe nanoparticles.

The results from compositional EDX-mapping on a sample grown using condition viii (Fig. 4) is in strong contrast to the one for condition iii in Fig 3. Here, one can clearly see the separate Fe- and Ni-particles generated, with no obvious intermixing of the elements. From this, it is clear that the nanoparticles of both materials are collected within the same nanotruss structure, hence forming nanotrusses built up of single-species Ni and Fe nanoparticles. Such structures could be of interest for applications where the interfaces between Ni and Fe are of interest [DDD], rather than alloys of the two materials. This collection of single-species nanoparticles in Fig. 4 is representative for the conditions vii-ix, even though the ratios of single element Fe nanoparticles and Ni nanoparticles are of course different for conditions vii and ix.

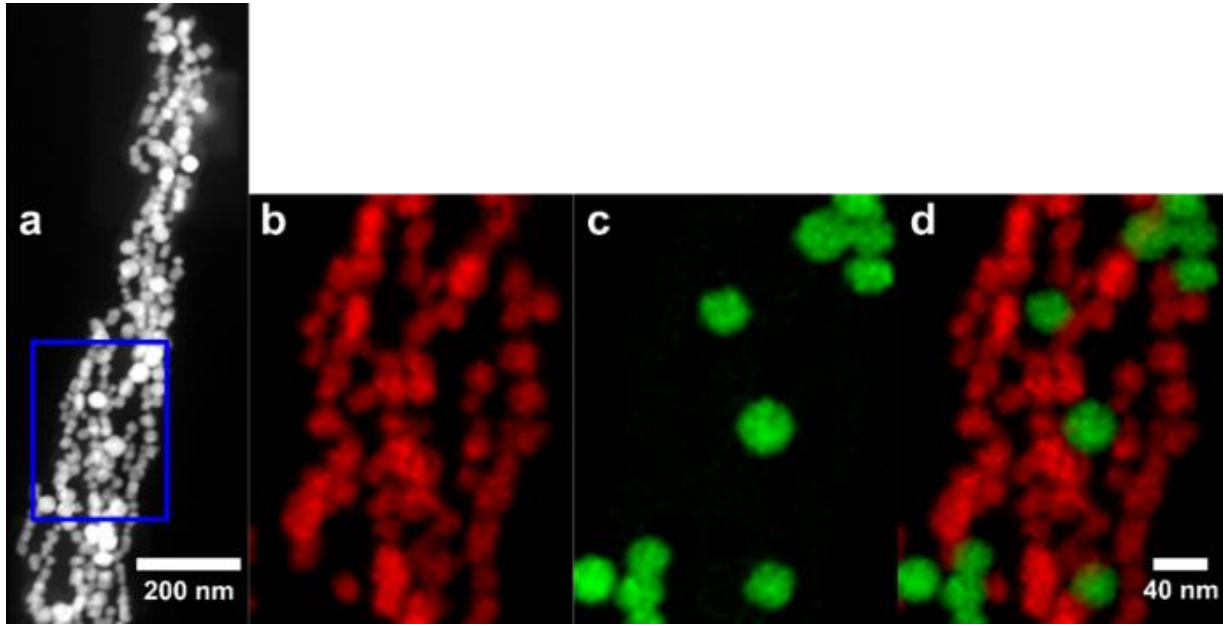

Fig. 4 | a) Dark field TEM image of a nanotruss grown using condition viii. b) and c) shows EDX mapping of the area marked in a) for Fe and Ni respectively, and d) is a combined image of b) and c).

XRD-analysis was performed on all sample conditions i-x. The results confirm that condition ii-iv consists of different crystal structures than the other conditions, as the main peaks are no longer α-Fe (110) or Ni (111), which is the case for all other conditions. For ii-iv, the peaks are closer to that of FeNi (111), with their exact peak position depending on the ratio of Fe/Ni. The results can be seen in the SI, Fig. S2, and further supports the conclusion that the structure grown using conditions ii-iv are alloys.

*Conclusion*

We have demonstrated a new type of sputter source for rapid growth of nanoparticles of two different elements. The particle composition is easy to tune thanks to the many adjustable parameters of the setup, as well as the rapid growth, which is one of the trademarks of the high-power hollow cathode nanoparticle sputtering.

**Nanowire design by magnetic collection of Fe, Ni and/or FeNi-alloy nanoparticles**

**Supplementary information**

S. Ekeroth[1], R. D. Boyd[1], N. Brenning[1,2], U. Helmersson[1]

[1] Department of Physics, Chemistry and Biology, Linköping University, SE-581 83 Linköping, Sweden.

[2] KTH Royal Institute of Technology, School of Electrical Engineering, Division of Space and Plasma Physics, SE-100 44 Stockholm, Sweden.

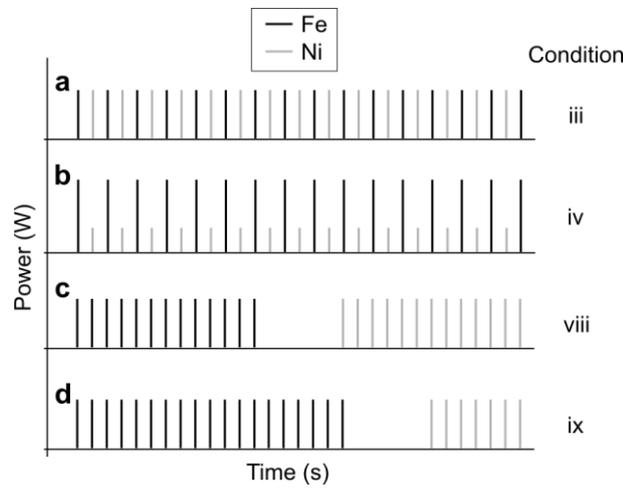

Fig. S1 | Schematic illustration of pulse delivery to the two cathodes.

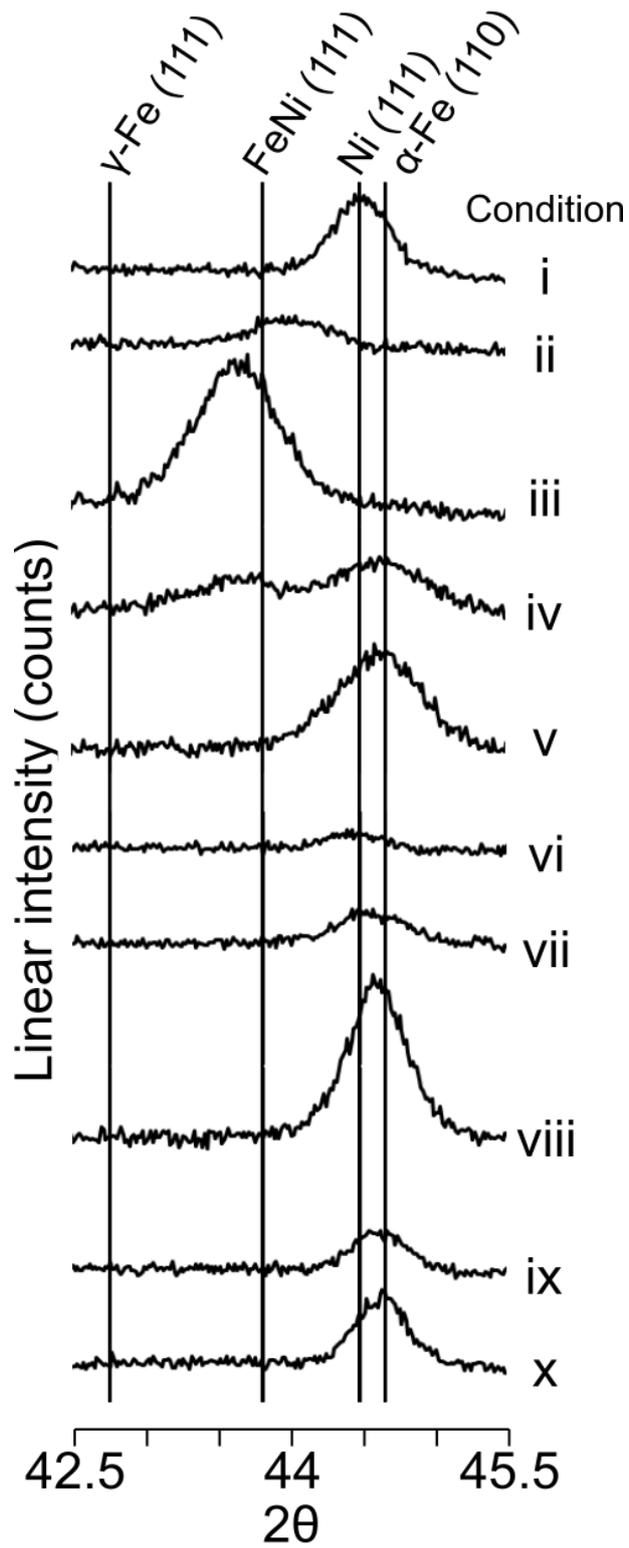

Fig. S2 | XRD diffractograms of nanoparticles grown using the conditions i-x.